# Superconductivity in Co-doped LaFeAsO

Athena S. Sefat♦, Ashfia Huq*, Michael A. McGuire♦, Rongying Jin♦,
Brian C. Sales♦, David Mandrus♦

*Materials Science & Technology Division♦ & Neutron Scattering Science Division**
*Oak Ridge National Laboratory, Oak Ridge, TN 37831, USA*

Lachlan M. D. Cranswick

*Canadian Neutron Beam Centre, National Research Council, Chalk River, ON K0J 1J0, Canada*

Peter W. Stephens, Kevin H. Stone

*Stony Brook University, Dept. of Phys. & Astron., Stony Brook, NY 11794, USA*

## Abstract

Here we report the synthesis and basic characterization of $LaFe_{1-x}Co_xAsO$ for several values of x. The parent phase LaFeAsO orders antiferromagnetically ($T_N \approx 145$ K). Replacing Fe with Co is expected to both electron dope the system and introduce disorder in the FeAs layer. For x = 0.05 antiferromagnetic order is destroyed and superconductivity is observed at $T_c^{onset} = 11.2$ K. For x = 0.11 superconductivity is observed at $T_c^{onset} = 14.3$ K, and for x = 0.15 $T_c^{onset} = 6.0$ K. Superconductivity is not observed for x = 0.2 and 0.5, but for x = 1, the material appears to be ferromagnetic ($T_c \approx$ 56 K) as judged by magnetization measurements. We conclude that Co is an effective dopant to induce superconductivity. Somewhat surprisingly, the system appears to tolerate considerable disorder in the FeAs planes.

## 1. Introduction

The recent reports of $T_c \sim 26$ K in $LaFeAsO_{1-x}F_x$ (x ~ 0.11) [1-10] and related materials have attracted a great deal of attention as these materials appear to be unconventional high-Tc superconductors that are based on iron instead of copper. It is interesting to compare and contrast the behavior of the new iron arsenide superconductors with the cuprates. For high-$T_c$ copper oxides, the parent compounds are antiferromagnetic Mott insulators [11], and the magnetic order is suppressed and superconductivity emerges by doping with either electrons or holes [12]. Much like copper oxide superconductors, high-$T_c$ in the Fe-based family of *R*FeAsO (*R* = rare-earth) is caused by doping. Here is the first report of superconductivity in LaFeAsO by electron-doping by means of cobalt substitution in the layers of $FeAs_4$ tetrahedra in $LaFe_{1-x}Co_xAsO$ (Fig. 1).

The parent LaFeAsO crystallizes with the ZrCuSiAs type structure [13], in *P*4/*nmm* space group (No. 129; Z = 2) [14, 15]. Electronic structure calculations [16] describe the structure as quasi two-dimensional, composed of sheets of metallic $Fe^{2+}$ in between ionic blocks of $LaOAs^{2-}$ along the *c*-axis. Although there is bonding between Fe and As (d = 2.3 - 2.4 Å), the states near the Fermi level are dominated by Fe *d*-states lightly mixed with As *p*-states. LaFeAsO undergoes a continuous or weakly first order structural phase transition from tetragonal (*P4/nmm*) to orthorhombic (*Cmma*) upon cooling below 155 - 160 K [8, 17, 18]; this is followed by a commensurate antiferromagnetic order just below 135-150 K [7, 8 19]. Ab initio calculations yield Fe magnetic moments ranging from 1.5 $\mu_B$ to 2.3 $\mu_B$ [5, 20-22], while weak superlattice reflections in neutron scattering [17], and Mössbauer spectra [7, 8] indicate a much smaller value of 0.25 - 0.35 $\mu_B$/Fe.

Carrier doping plays a major role in the appearance of superconductivity, by suppressing the magnetic order and structural phase transition in $LaFeAsO_{1-x}F_x$ [7, 23, 24]. Substitution of La for other rare-earths (*R*) has resulted in finding F-doped $R$FeAs$O_{1-x}F_x$ (*R* = Pr, Sm, Nd, Gd) superconductors [25-30], giving $T_c$ as high as 55 K in $SmFeAsO_{0.9}F_{0.1}$ [30]. $Th^{4+}$-doping in $Gd_{1-x}Th_xFeAsO$ gives slightly higher $T_c$ at 56 K [31]. Hole doping, by substituting $Sr^{2+}$ for $La^{3+}$ results in $T_c$ = 25 K [32]. A series of oxygen deficient $R$FeAs$O_{1-\delta}$ have also been prepared by high pressure syntheses with highest $T_c$ = 55 K for $SmFeAsO_{0.85}$ [33-35]. Here we report of the synthesis and characterization of Co-doped LaFeAsO, and find superconductivity for ~ 4 - 12% doping levels. Chemically, Co is a better means of electron doping, as carriers are doped directly into the FeAs planes. This also provides information as to how well the new superconductors tolerate in-plane disorder.

For $LaFe_{1-x}Co_xAsO$ ($0 \le x \le 1$), experimental details below are followed by a discussion of the crystal structure from powder x-ray and neutron diffraction. The thermodynamic and transport properties of this material will then be presented and discussed. The measurements include field and temperature-dependent magnetic susceptibility, electrical resistivity, and Seebeck coefficient.

## 2. Experimental details

Polycrystalline samples with $LaFe_{1-x}Co_xAsO$ (x = 0, 0.05, 0.11, 0.15, 0.2, 0.5, 1) nominal compositions were synthesized by stoichiometrically mixing fine powders of LaAs [2], $Co_3O_4$ (99.9985%), $Fe_2O_3$ (99.99 %, calcined at 900 °C for 12 hrs), and Fe (99.998 %), pressing into a pellet, and rapidly heating in silica tubes. The silica tubes were sealed under partial pressure of argon and heated at 1220 °C for ~ 12 hrs, then rapidly cooled by shutting off the furnace. Each pellet was reground and re-annealed. The source of all elements or compounds was Alfa Aesar.

The initial phase purity and structural identification were made via powder x-ray diffraction, using a Scintag XDS 2000 θ-θ diffractometer (Cu $K_\alpha$ radiation). The cell parameters for LaCoAsO was refined using least squares fitting of the measured peak positions in the 2θ = 20 - 70° using Jade 6.1 MDI package.

Neutron powder diffraction data on $LaFe_{1-x}Co_xAsO$, with x = 0, 0.11, and 0.15 were collected on the C2 diffractometer [36] operated by the Neutron Program for Materials Research of the National Research Council of Canada. Each sample of 1 - 2 g was placed in a helium-filled vanadium can that was sealed with an indium gasket. Data were collected at two different wavelengths of 1.330910 Å and 2.372630 Å. For x = 0 and 0.15, powder diffraction data were collected at 4 K and also 300 K. For x = 0.11, data were collected at 4, 10, 15, 30, 50, 100, 200, and 300 K. Neutron powder diffraction data for both wavelengths were jointly refined for the main phase and impurity phases of FeAs, and $La_2O_3$, using GSAS [37, 38] with the EXPGUI [39] interface. For all the refinements, the pseudo-Voigt function (profile function 3 in GSAS) was used and both Lorentzian and Gaussian widths were allowed to vary. Zero point parameter was also allowed to change. The background was modeled using a 10 term Chebyshev polynomial function. The lattice parameters for the main as well as impurity phases were refined. Isotropic thermal parameters and fractional coordinates were refined for all the atoms in the primary phase. For x = 0.11 and 0.15 samples, a single thermal parameter was used for both Fe and Co and the occupancy factors were refined with the constraint that they add up to the full occupancy of the site. The details of the neutron powder refinement conditions and the atomic parameters, at 4 K or 300 K for *d* range of 0.8 to 10.6 Å, are given in Table 1 and 2, for x = 0 and 0.11, respectively.

Powder x-ray diffraction patterns were collected at beam-line X16C of the National Synchrotron Light Source utilizing x-rays of wavelength 0.69869 Å from a Si(111) channel-cut monochromator and a Ge(111) analyzer. The data for $LaFe_{1-x}Co_xAsO$ with x = 0.05, 0.11, 0.2, and 0.5 were collected at room temperature. Samples were ground in air and diluted with ground glass wool, and flame-sealed in 1 mm capillaries. Total exposure to the atmosphere before sealing was approximately one minute for each sample. Capillaries were spun during data collection to improve counting statistics. The high resolution x-ray data were modeled in a similar fashion as the neutron data described above.

DC magnetization was measured as a function of temperature and field using a Quantum Design magnetic property measurement system (MPMS). For a temperature sweep experiment, the sample was cooled to 1.8 K in zero-field (zfc) and data were collected by warming from 1.8 K to 300 K in an applied field. The sample was then cooled in the applied field (fc), and the measurement repeated from 1.8 K. The magnetic susceptibility results are presented per mole of $LaFe_{1-x}Co_xAsO$ ($0 \le x \le 1$) formula unit ($cm^3$/mol). Electrical resistivity, Seebeck coefficient, and specific heat measurements were obtained using a Quantum Design physical property measurement system (PPMS). For dc resistance measurements, electrical contacts were placed on samples in standard 4-probe geometry, using Pt wires and silver epoxy (EPO-TEK H20E). Gold-coated copper leads were used for Seebeck coefficient measurements.

## 3. Results and Discussion

### 3.1 Structure

For $LaFe_{1-x}Co_xAsO$, neutron diffraction data were collected on x = 0, 0.11, 0.15, while x-ray diffractions were done on x = 0.05, 0.11, 0.2, and 0.5 compositions.

The structure of $LaFe_{1-x}Co_xAsO$ at 300 K for all compositions of x, was refined with the tetragonal space group *P*4/*nmm* (No. 129, origin choice 2; Z = 2) [14, 15] (Fig. 1). La and As atoms are located at Wyckoff positions 2*c*, O site is at 2*a*, while Fe/Co may be shared at site 2*b*. For LaFeAsO, the lattice constants obtained from neutron data are $a$ = 4.0345(1) Å and $c$ = 8.7387(4) Å, and compare well with the reported $a$ = 4.038(1) Å and $c$ = 8.753(6) Å [15]. The incorporation of Co in x= 0.11 in the Fe site reduces the cell volume by 0.26 %, due to contraction of the lattice constant $c$ (= 8.7132(3) Å). For F-doped $LaFeAsO_{1-x}F_x$ with x = 0.11, the lattice constants are found to shrink by 0.17 % (4.0277(2) Å) and 0.30% (8.7125(4) Å), respectively [2].

The structure of LaFeAsO at 4 K was refined with orthorhombic space group C*mma* (No. 67; Z = 4), first described by T. Nomura et al. and refined using x-ray synchrotron diffraction data at 120 K [18]. As a result of this phase transition, the *a*- and *b*-crystallographic axes rotate by 45°. Splitting of several reflections of the room temperature tetragonal phase (111, 112, 322) was observed as the temperature was lowered to 4 K. The atoms in this low temperature orthorhombic phase are characterized by the same coordination as that of tetragonal structure. La and As atoms are located on Wyckoff positions 4*g*, Fe on 4*b*, and O on 4*a*.

For $LaFe_{1-x}Co_xAsO$ with x = 0.11, similar peak splitting to LaFeAsO was however not observed in the diffraction data, down to 4 K. The data was refined within the tetragonal *P*4/*nmm* framework, comparable to that for 14 % F-doped LaFeAsO structure at 120 K [18]. The refined neutron diffraction pattern at 4 K is shown in Fig. 2a. The data were refined for the main phase (89.8 %) and impurity phases of FeAs (4.1 %), and $La_2O_3$ (6.1 %).

For x = 0.11, the temperature dependence of lattice parameters and the cell volumes are shown in Fig. 2b. The *a*- and *c*-parameters seem to vary nonlinearly with increasing temperature and by 0.18 % and 0.31 %, respectively, over the 4 K to 300 K region; the cell volume change is 0.68 %. It has been suggested that the *R*-As distance and Fe-As-Fe angle are the two crucial parameters in the structure of *R*FeAsO that control the effective band widths in these materials and hence the superconductivity [40]. At room temperature, the variation of the bond distance and angle of the x = 0.11 versus Fe-parent [41] is very small (La-As: 0.06 %; Fe-As-Fe: 0.05 %). However, these differences are significantly larger at 4 K (La-As: 0.11 %; Fe-As-Fe: -0.14 %). It is interesting to note that the large distortion of the $FeAs_4$ tetrahedron in x = 0.11 is accompanied by a similarly large distortion in $La_4O$ tetrahedron, whereas F-doping in the La-O layer in LaFeAsO, causes no distortion in $FeAs_4$ tetrahedron. [40]. F-doping in the $La_4O$ layer changes the Fe-As bond only slightly, however, Co doping in the $FeAs_4$ layer changes the La-O bond significantly.

The occupancy of Co on Fe 2*b* site refines as $LaFe_{0.92(1)}Co_{0.08(1)}AsO$ for x = 0.11. Also, the average occupancy of Co for nominally prepared x = 0.15 sample refines as x = 0.12(1). Assuming a linear dependence between nominal x and the Co concentration, the

x = 0.05, 0.2 and 0.5 samples are likely to have ~ 4 %, 16 %, and 40 % cobalt. Despite the variance between the nominal and measured compositions, they will be presented with their nominal x values. The three compositions are stoichiometric in oxygen content as 2*a* site was refined to unity at every temperature that the data were collected.

Lattice parameters and cell volume, as a function of cobalt doping at 300 K, were determined from x-ray diffraction data (Fig. 3). From x = 0 to x = 1, the reduction of the *a*-parameter, *c*-parameter, and cell volumes are 0.04 %, 3.20 %, and 2.79 %, respectively. It was suggested that the $T_c$ in $R$FeAs$O_{1-y}$ is strongly dependent on the tetrahedral distortion represented by As-Fe-As bond angle [42]. We also observe similar trend from out limited x-ray data at various concentration of Co. From x = 0.11 to 0.2, As-Fe-As bond angle decreases from 107.35° to 106.99° [41]. It should be mentioned here that at x = 0.2, superconductivity is lost (Sec. 3.2).

### 3.2 Physical properties

The temperature dependence of molar magnetic susceptibility in 1 Tesla for the $LaFe_{1-x}Co_xAsO$ system is shown in Fig. 4. For $0 \leq x \leq 0.5$, $\chi$ values vary between $3 \times 10^{-5}$ and $3.5 \times 10^{-4}$ $cm^3$/mol at 300 K. These values are smaller than that reported for $LaFeAsO_{0.89}F_{0.11}$ ($1.8 \times 10^{-3}$ $cm^3$/mol, at 300 K and 0.1 T) [2]. For LaFeAsO (Fig. 5), magnetic susceptibility decreases with decreasing temperature, with an anomaly at ~ 150 K, and a rise below 50 K. For x = 0.2 and 0.5, $\chi$ is weakly temperature dependent, with gradual increase below 50 K. The field dependent magnetization data for x = 0, 0.2 and 0.5 are shown in the inset of Fig. 4 and none of the samples approach saturation at 7 T, reaching < 0.003 $\mu_B$/f.u.

The temperature dependence of magnetization in 1 Tesla for LaCoAsO is shown in Fig. 5a. The M(T) increases below ~ 80 K, indicating possible ferromagnetic ordering. The inset of Fig. 5a shows a plot of the inverse magnetization and the fit to a Curie-Weiss law, where $M/H \approx \chi = C / (T-\theta) + \chi_o$; $C$ is the Curie constant, $\theta$ is the paramagnetic Weiss temperature, and $\chi_o$ is a temperature-independent term. The fit was done for T > 150 K. Assuming Co is the only moment bearing ion, then the term $C$ is related to the effective moment in paramagnetic state and the theoretically expected 3.87 $\mu_B$ ($S$ = 3/2). Experimentally, the Curie-Weiss fit gives $\mu_{eff} \approx 1.4$ $\mu_B$ per Co. This result is comparable to that recently reported for LaCoAsO [43]. The Weiss temperature is positive and $\theta \approx 90$ K; $\chi_o \approx -6 \times 10^{-6}$/mol. Field-dependent magnetization data, M(H), at 1.8 K are shown in Fig. 5b. The magnetization rises rapidly up to ~ 500 G and changes weakly thereafter. The precipitous rise in magnetization at low fields again suggests ferromagnetism, although as noted below, there is no anomaly observed in specific heat. An applied field of ~ 5 T gives 0.46 $\mu_B$ per formula unit; this field is insufficient to saturate the magnetization data to 3 $\mu_B/Co^{2+}$, as expected for localized high-spin $Co^{2+}$ ions.

In order to locate the ferromagnetic transition, magnetization isotherms in the vicinity of Curie temperatures have been measured for the construction of Arrott plots [44] in the form of $M^2$ versus $HM^{-1}$ (Fig. 5b, inset); for this sample, 55 K < $T_c$ < 58 K. This range of $T_c$ is slightly lower than that reported earlier [1, 43]. Despite the suggestion of a ferromagnetic component in magnetization measurements, there is lack of anomaly in specific heat on several different polycrystalline samples. One scenario that may explain the data is that LaCoAsO is a quasi-2D itinerant metal, with strong in-plane ferromagnetic interactions but extremely weak out-of-plane interactions. In this scenario, LaCoAsO would be a strongly exchange-enhanced paramagnet, and that application of even a small field produces a large ferromagnetic response.

Fig. 6 shows the temperature dependence of magnetic susceptibility, measured under zfc and fc conditions at 20 Oe, for $LaFe_{1-x}Co_xAsO$ with $x \leq 0.15$. The susceptibility becomes negative below ~ 11 K, 14 K, and 6 K, for x = 0.05, 0.11, and 0.15, respectively. Assuming theoretical density of roughly 6.68 g/cm$^3$ and $\chi$ value of the perfect diamagnetism, we estimate the shielding fraction about 15% and Meissner fraction near 6% at 2 K for x = 0.05. Although it has higher $T_c$ value and higher shielding fraction (90%), the Meissner fraction is only about 2% for x = 0.11. For x = 0.15, both the shielding and Meissner fractions are less than 1%. It should be noted that since the Meissner fraction is determined by pinning and penetration effects, its interpretation is ambiguous on sintered samples.

Fig. 7a shows the temperature dependence of the electrical resistivity ($\rho$) in zero field. Similar to the other reports for LaFeAsO [5, 8], $\rho_{300\,K} \approx 4$ m$\Omega$ cm, then it increases slightly with decreasing temperature, peaking at ~ 160 K. This upturn in $\rho$ is likely associated with the increased charge carrier scattering by lattice fluctuations related to the onset of the structural transition. The samples with x = 0.05, 0.11, and 0.15 exhibit superconductivity. Higher doping levels of x = 0.2, 0.5, and 1 give only metallic behavior. The electrical conductivity increases by a factor of eight, from LaFeAsO ($3d^6$) to LaCoAsO ($3d^7$).

The resistive transitions for the superconducting compositions shift to lower temperatures by applying a magnetic field (Fig. 7b). The transition width for each sample becomes wider with increasing H, a characteristic of type-II superconductivity. Here we define a transition temperature $T_c(H)$ which satisfies the condition that $\rho(T_c, H)$ equals a fixed percentage of the normal-state value ($\rho_N$) for each field H. The $T_c(H)$ values for $\rho$ = 10, 50, and 90 % are shown as insets of Fig. 7b, represented by the upper critical field $H_{c2}(T)$. In all cases we find that $H_{c2}(T)$ has a linear dependence with no sign of saturation. $T_c$ at zero field for $0.1\rho_N$ are 11.2 K, 14.3 K, and 6.0 K, for x = 0.05, 0.11, and 0.15, respectively. The transition width $\Delta T_c = T_c(90\%) - T_c(10\%)$ are 3.2 K, 2.3 K, and 3.9 K, respectively. The $\Delta T_c$ values are smaller than the reported 4.5 K for $LaFeAsO_{0.89}F_{0.11}$ with $T_c^{onset}$ = 28.2 K [2].

Fig. 8a shows the temperature dependence of specific heat in the form of C/T versus $T^2$, from 1.8 K to 20 K. For each x = 0.05 and 0.11, there is clearly a broad specific heat anomaly below ~ $T_c$. For x = 0.2, 0.5, and 1 samples, the C/T vs $T^2$ plot is linear between ~ 2 and 6 K (Fig. 8a, bottom). This allows the estimation of electronic $\gamma$ and lattice $\beta$ values, as $C = \gamma T + \beta T^3$. The fits yield $\gamma$, in units of mJ/(K$^2$ mol atom) for x = 0.2, 0.5, and 1, respectively, as 1.405(1), 1.68(1), and 4.02(1). The value of the Debye temperature ($\theta_D$) was subsequently calculated in the low temperature limit ($\beta = 12\pi^4 R/5\theta^3$), and $\theta_D$ = 342(9) K for these samples. This value is comparable to the Debye temperature of ≈ 280 K for LaFeAsO [5] and ≈ 325 K for $LaFeAsO_{0.89}F_{0.11}$ [2].

For x= 0.05, the specific heat under magnetic fields of 14 T was measured and subtracted from zero-field data. Fig. 8b gives the temperature dependence of $\Delta C/T$ data and illustrates a deviation from high-temperature behavior below $T_c$ ~ 11 K, peaking at 7

K. The broadened transition reflects the inhomogenity of superconducting phase, likely due to the random Co distribution. Nevertheless, the observation of specific heat anomaly at $T_c$ indicates bulk superconductivity in the newly discovered Co doped LaFeAsO. The origin of the field dependence of the background is not understood and has yet to be investigated.

The Seebeck coefficient, S, of $LaFe_{1-x}Co_xAsO$ with x = 0.11 at 0 T is shown in the inset of Fig. 9. S is negative over the entire temperature range, indicating dominant electron conduction. This result is similar to F-doping (x = 0.11) in LaFeAsO and provides evidence for electron doping [2]. S varies from -50 μV/K at 300 K to a value of ~ -65 μV/K at ~ 160 K, then decreases in magnitude as the temperature is lowered further. The maximum in the data is probably due to the competition between dominant electron-like bands and the expected proximity of hole-like bands near the Fermi energy. Fig. 9 gives the magnetic field effect below 25 K; $T_c$ at ~ 14 K is clearly suppressed at 8 T.

## 4. Conclusions

In this work we report the synthesis, structure, magnetization, resistivity, specific heat, and Seebeck coefficient of $LaFe_{1-x}Co_xAsO$ for several values of x. The most important observation is that Co acts as an effective dopant and produces superconductivity in this system. In most respects, $LaFe_{1-x}Co_xAsO$ behaves similarly to $LaFeAsO_{1-x}F_x$, but with a smaller $T_c$. This is likely due to the stronger effects of disorder produced by doping in the FeAs layers, rather than in the LaO layers. It is actually somewhat surprising that the superconductivity in $LaFe_{1-x}Co_xAsO$ is quite robust to in-plane disorder, and this behavior will need to be understood as part of a comprehensive theory of the superconducting mechanism.

## Acknowledgement

We would like to thank D. J. Singh for helpful discussions. The research at ORNL was sponsored by the Division of Materials Sciences and Engineering, Office of Basic Energy Sciences, U.S. Department of Energy. Part of this research was performed by Eugene P. Wigner Fellows at ORNL, managed by UT-Battelle, LLC, for the U.S. DOE under Contract DE-AC05-00OR22725. LMDC would like to thank CNBC NRC technicians, R. Sammon, D. Dean, and T. Dodd, for setup of the closed cycle refrigeration system used for the neutron diffraction measurements.

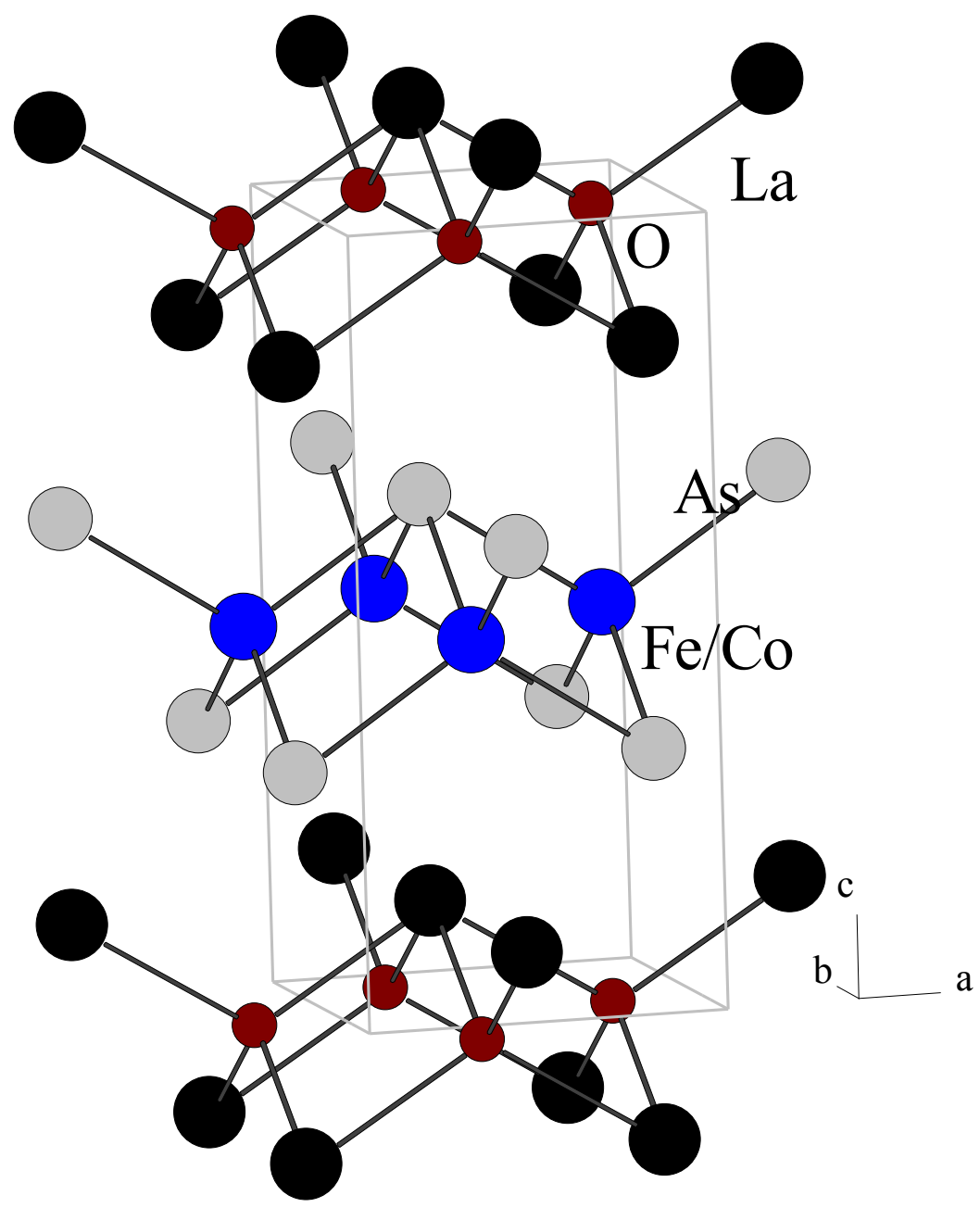

Fig. 1. Crystal structure of $LaFe_{1-x}Co_xAsO$, with tetragonal ZrCuSiAs-type. It is composed of layers of edge-sharing $OLa_4$-tetrahedra alternating with layers of $Fe(/Co)As_4$ along the *c*-axis. The unit cell is represented in grey lines.

(a)

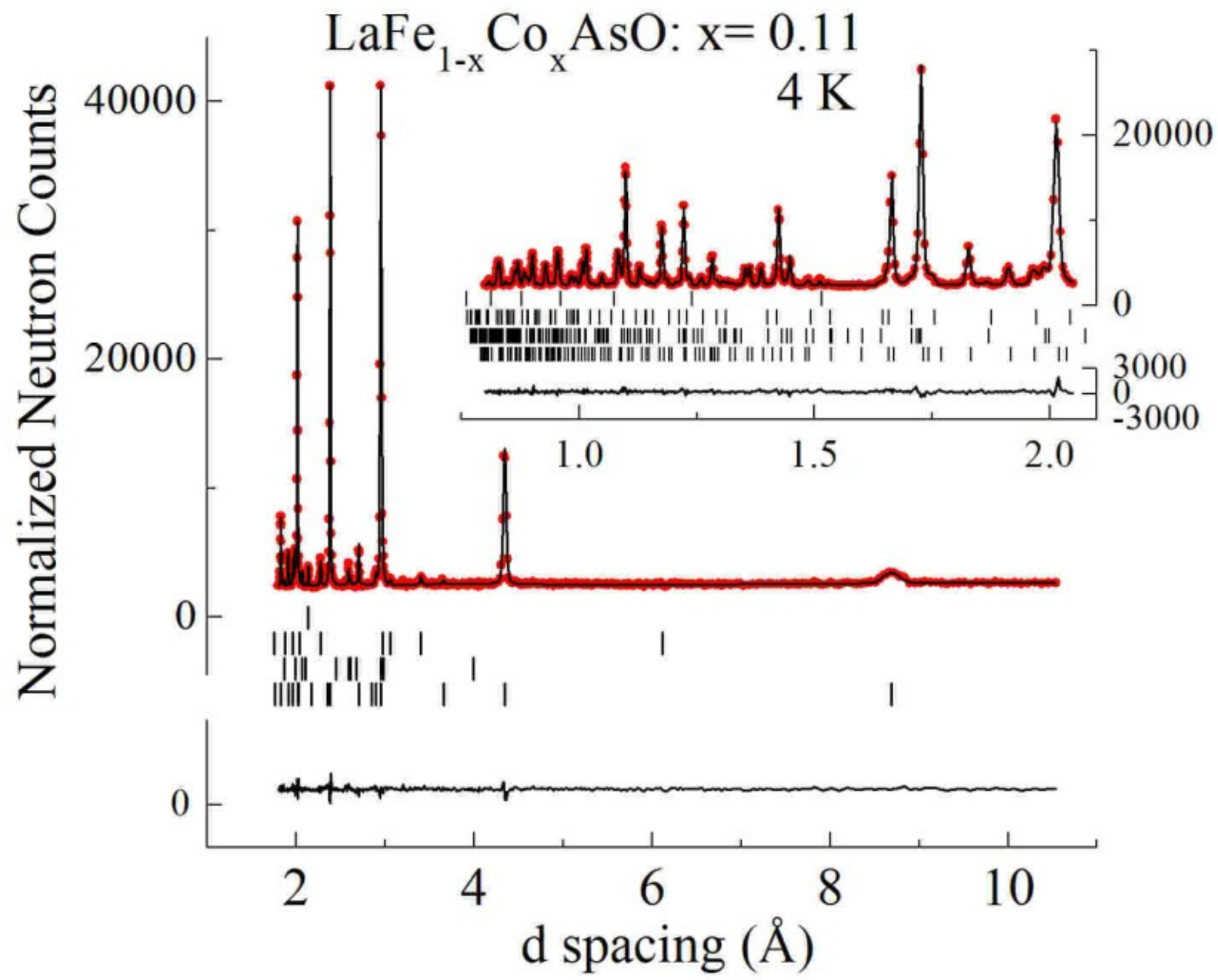

(b)

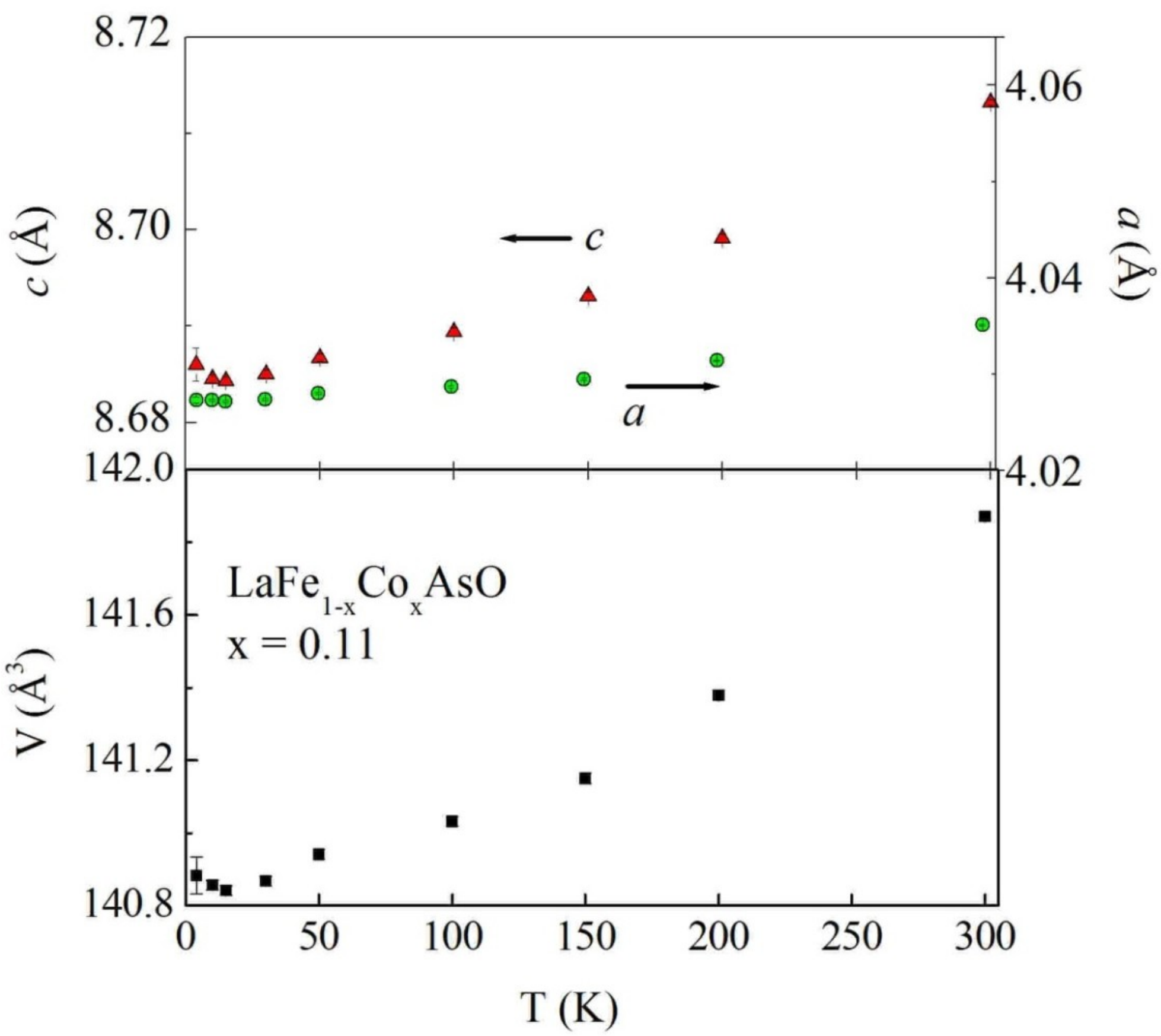

Fig. 2. (Color online) For $LaAsFe_{1-x}Co_xO$, x = 0.11, (a) the refined neutron powder diffraction profile at 4 K, and (b) the change of lattice parameters and cell volume with temperature. The tick marks in (a) are bottom to top: main phase in *P*4/*nmm*, FeAs, $La_2O_3$, and V.

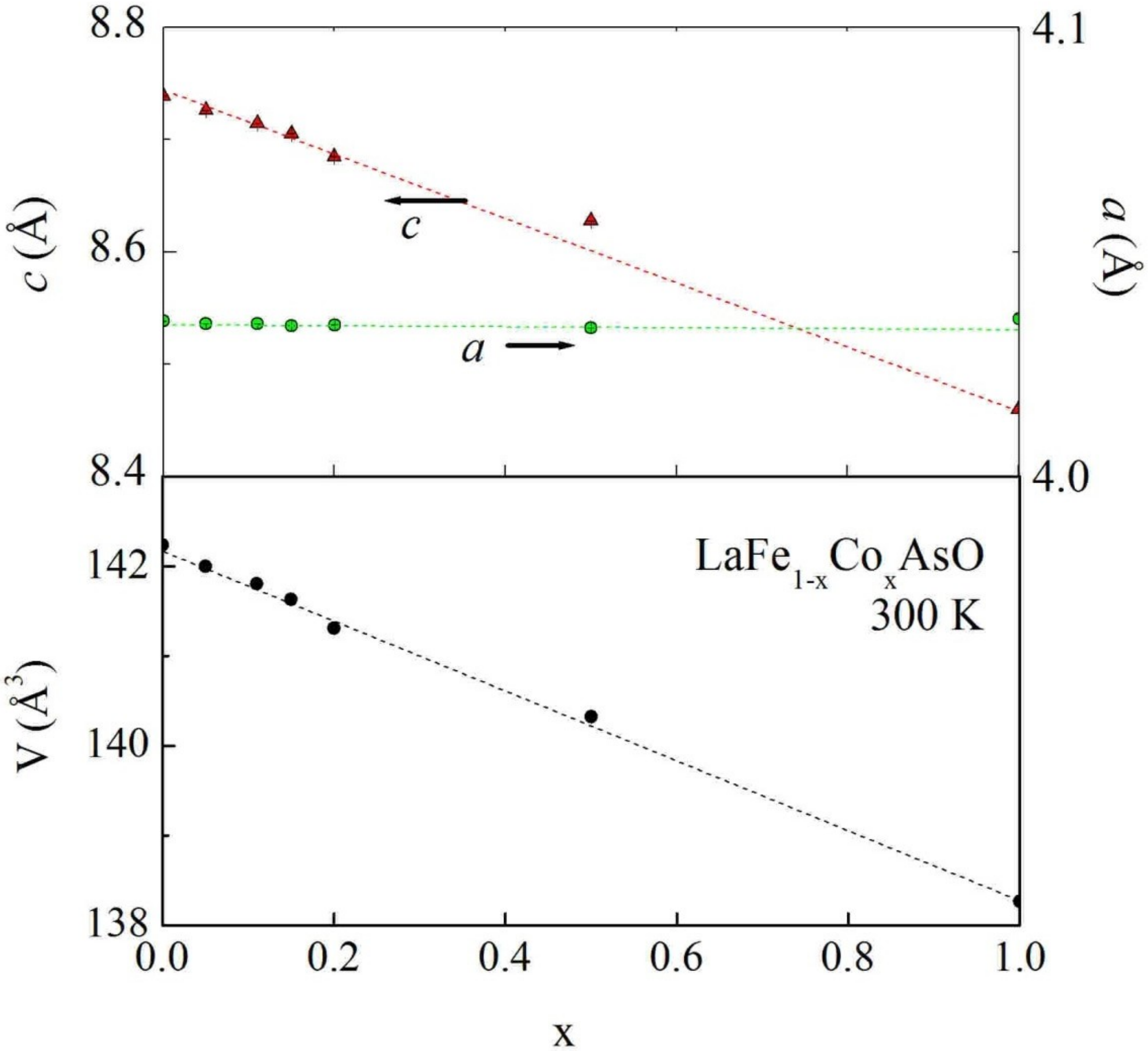

Fig. 3. (Color online) Refined lattice parameters and cell volume for $LaAsFe_{1-x}Co_xO$, $0 \leq x \leq 1$, at 300 K.

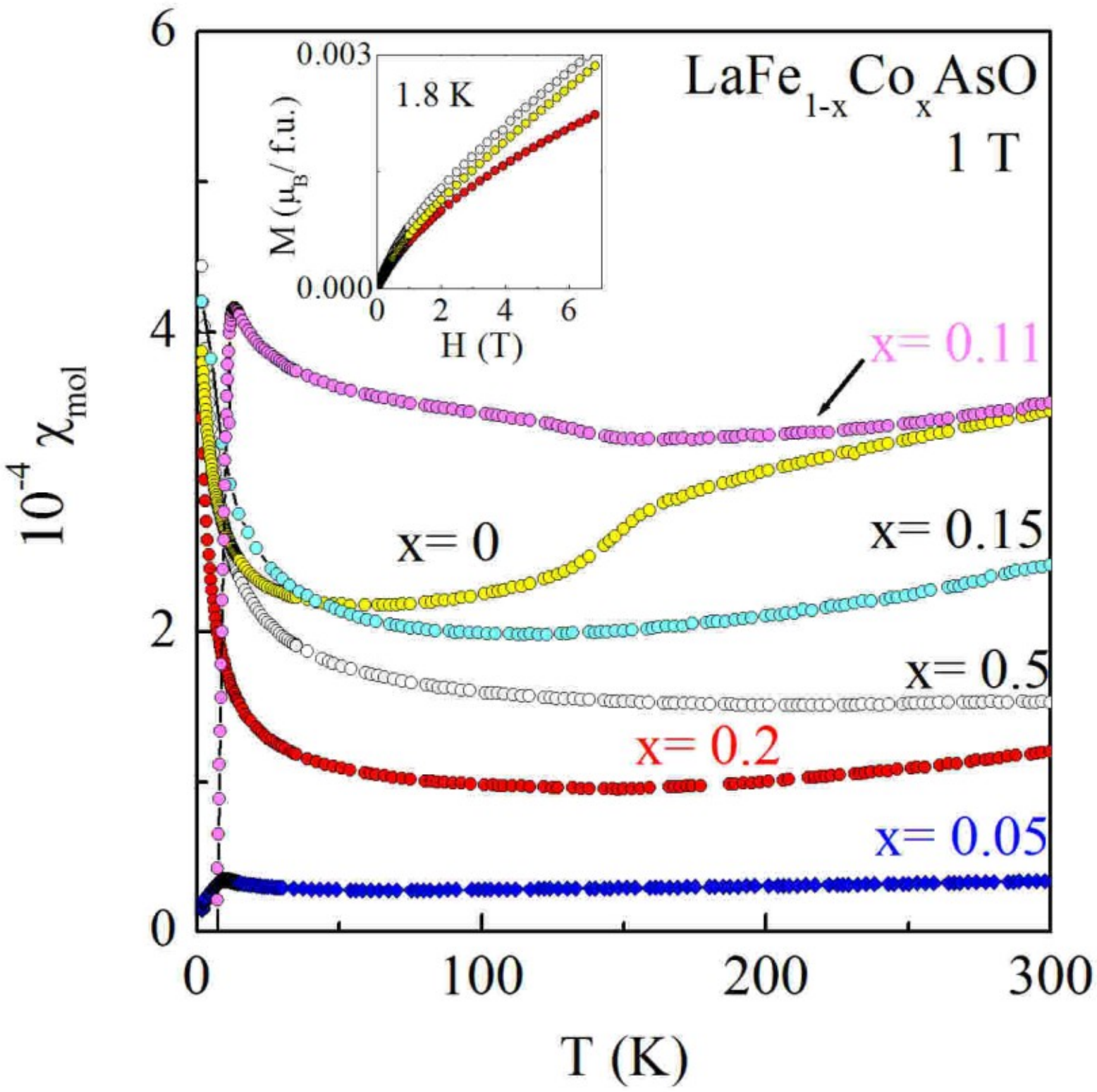

Fig. 4. The dependence of molar susceptibility in zero-field in $LaAsFe_{1-x}Co_xO$ for $0 \leq x \leq 0.5$. The field dependent magnetization at 1.8 K, for the non-superconducting compositions (inset).

(a) (b)

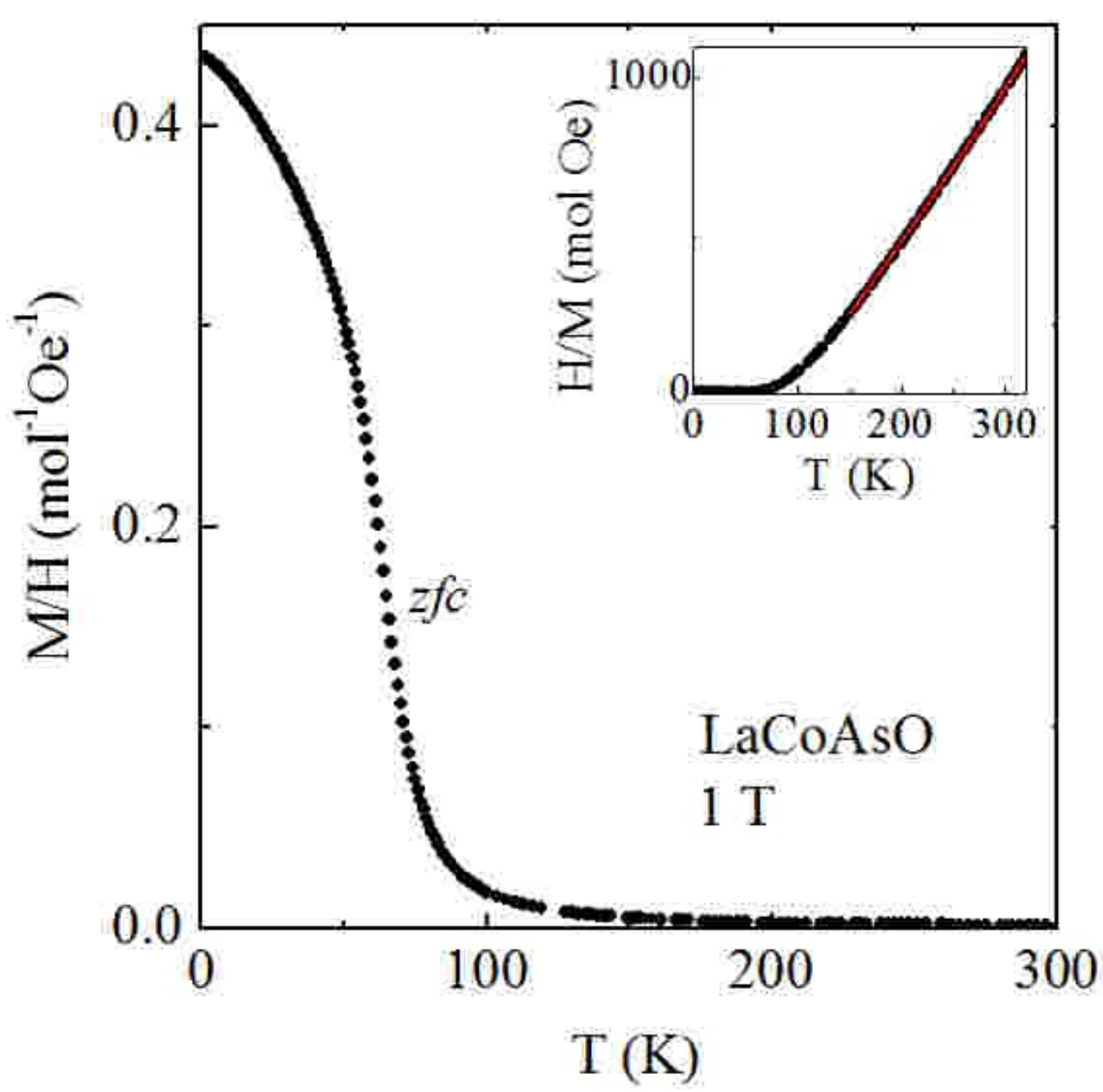

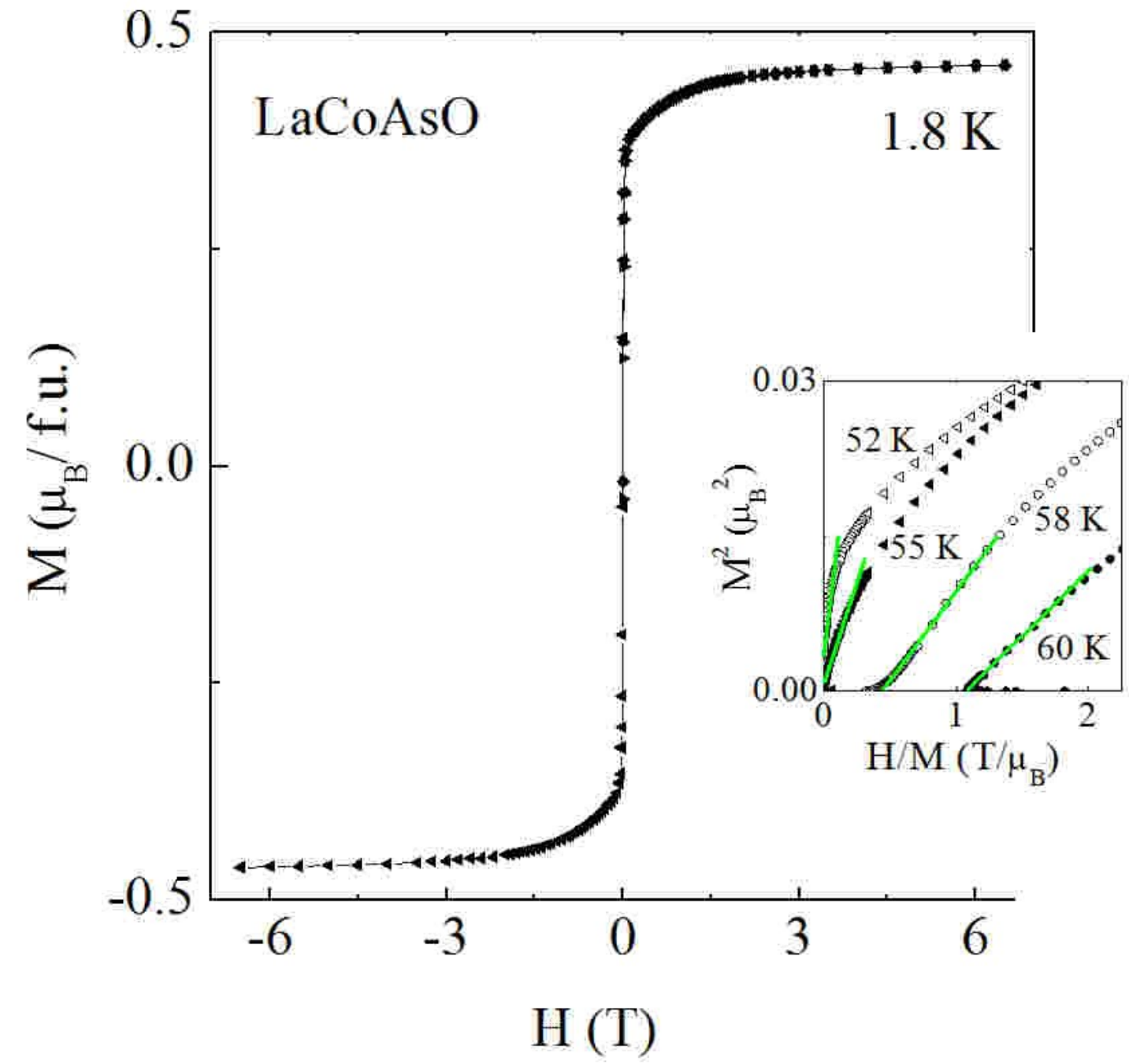

Fig. 5. (Color online) (a) Temperature dependence of molar susceptibility, at 1 T for LaCoAsO. The plot of the inverse molar susceptibility, with fit to modified Curie-Weiss above 150 K in red (inset). (b) Magnetization versus applied field at 1.8 K for LaCoAsO. The Arrott plots in the form of $M^2$ versus H/M, at temperatures in the vicinity of magnetic transition (inset).

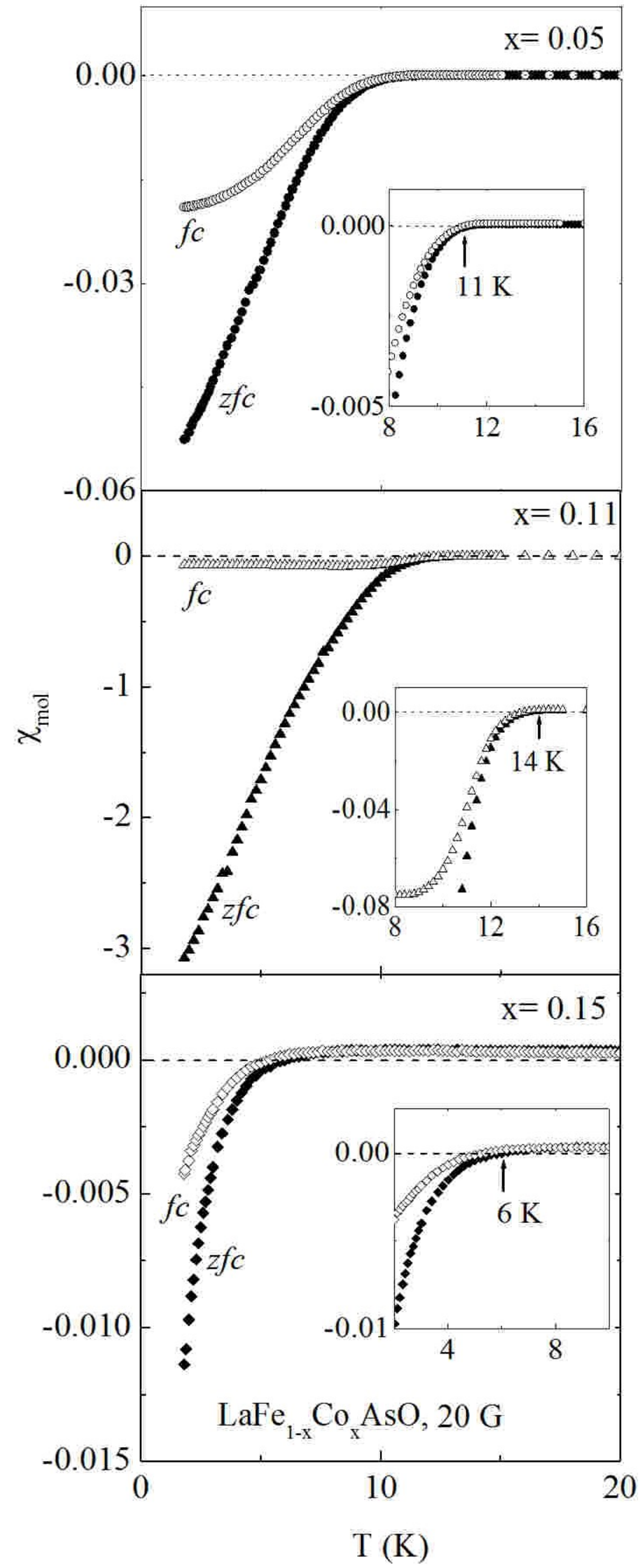

Fig. 6. Temperature dependence of molar susceptibility in zero-field cooled, zfc (filled symbols), and field-cooled, fc (open symbols), for $LaFe_{1-x}Co_xAsO$ in 20 Oe for x= 0.05 (top), x = 0.11 (middle), and 0.15 (bottom). The enlarged regions around $T_c$ are shown in the insets.

(a)

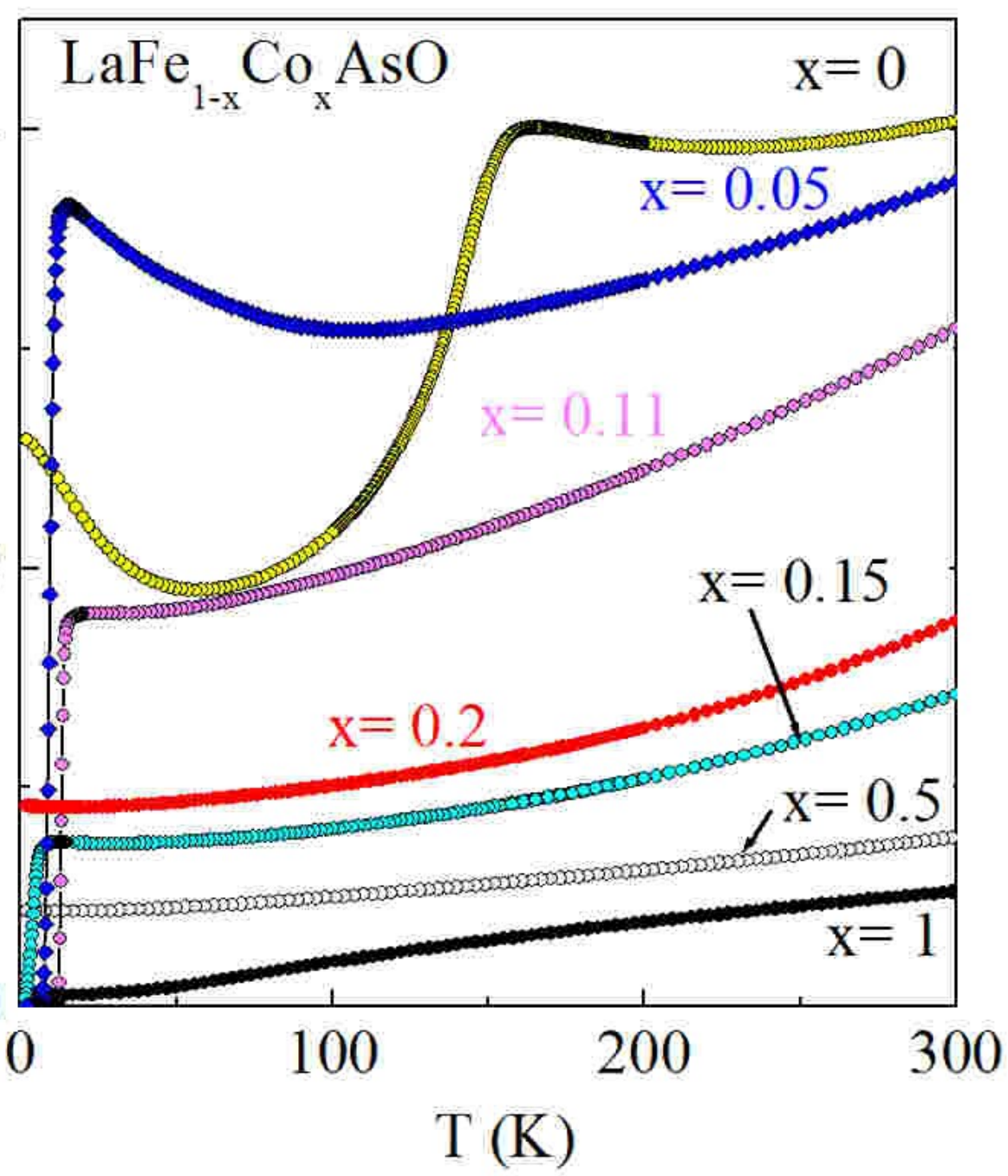

(b)

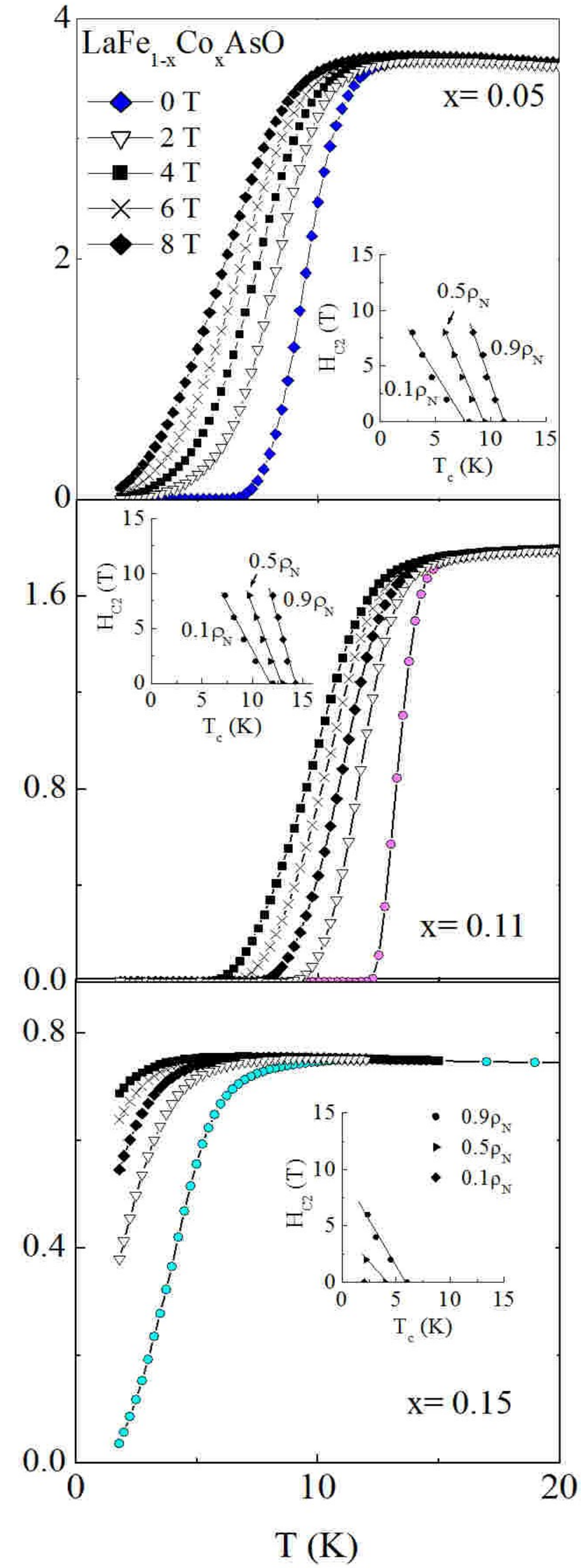

Fig. 7. (Color online) (a) Temperature dependence of resistivity for $LaFe_{1-x}Co_xAsO$ for $0 \leq x \leq 1$. (b) The temperature dependence of resistivity at various applied fields for x = 0.05 (top), 0.11 (middle), and 0.15 (bottom). For each composition, the upper critical field $H_{c2}$ is found from 90%, 50 % and 10% estimates of the normal-state value, $\rho_N$, and plotted versus critical temperature as insets

(a)

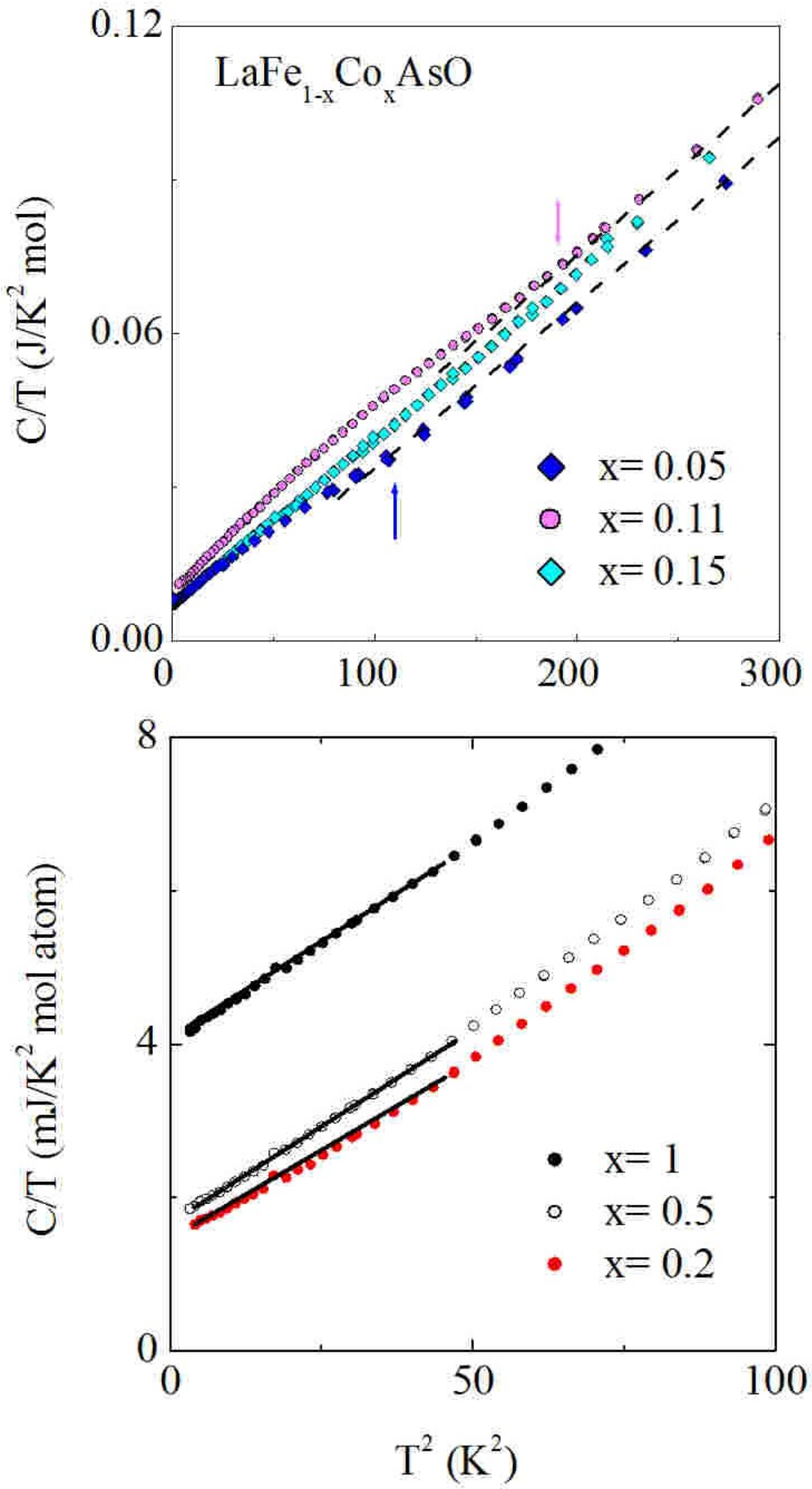

(b)

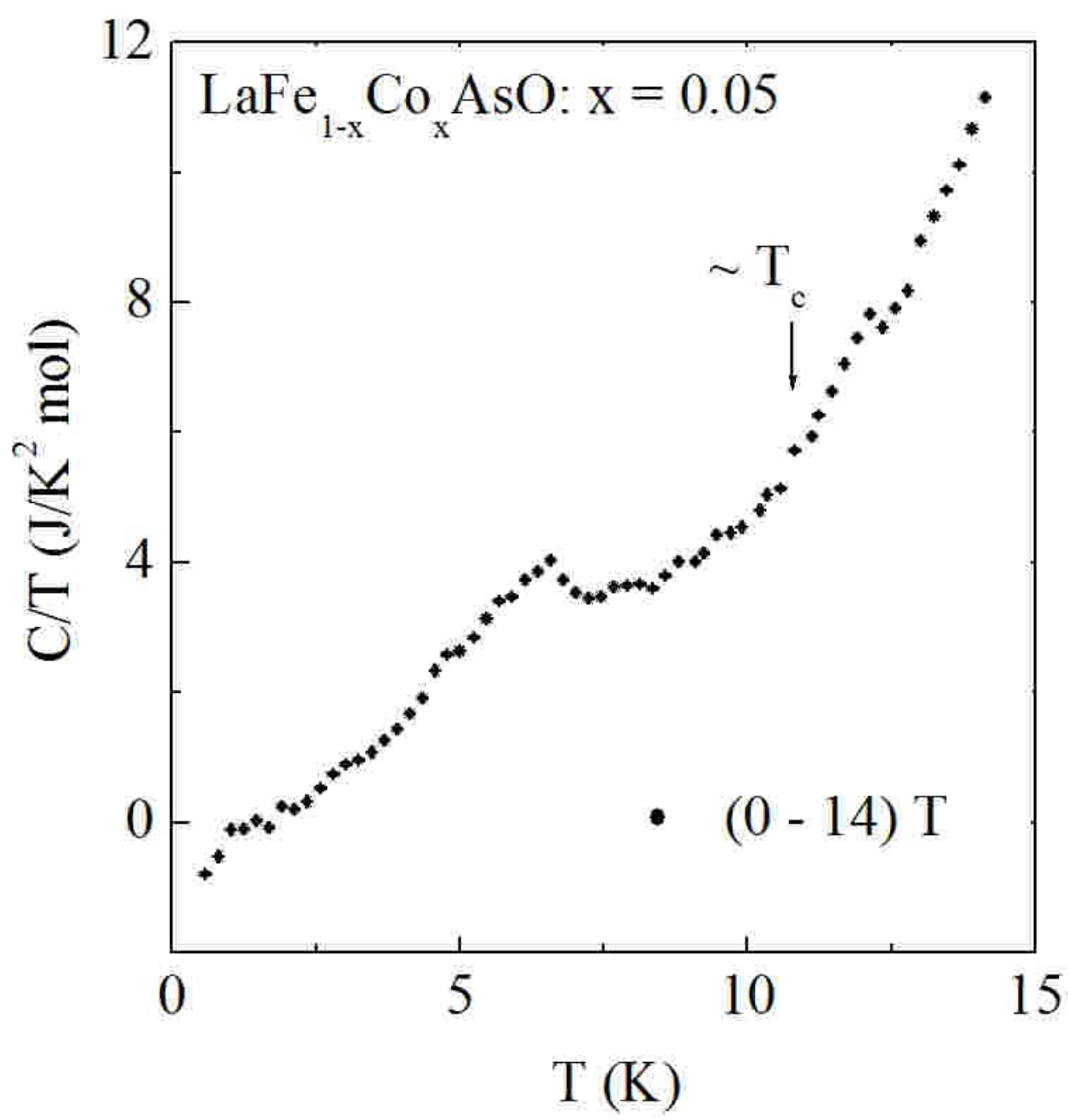

Fig. 8. (Color online) (a) For $LaFe_{1-x}Co_xAsO$ and $0.05 \le x \le 1$, temperature dependence of specific heat in the form of C/T versus $T^2$, below ~ 17 K (top) and 10 K (bottom). Anomalies are shown by arrows for x = 0.05 and 0.11. For $0.2 \le x \le 1$ (bottom), linear fits are below ~ 6 K. (b) For $LaFe_{1-x}Co_xAsO$ and x = 0.05, the temperature dependence of the subtracted 0 T from 14 T specific heat data. The arrow shows the onset of the deviation from the background at ~ $T_c$.

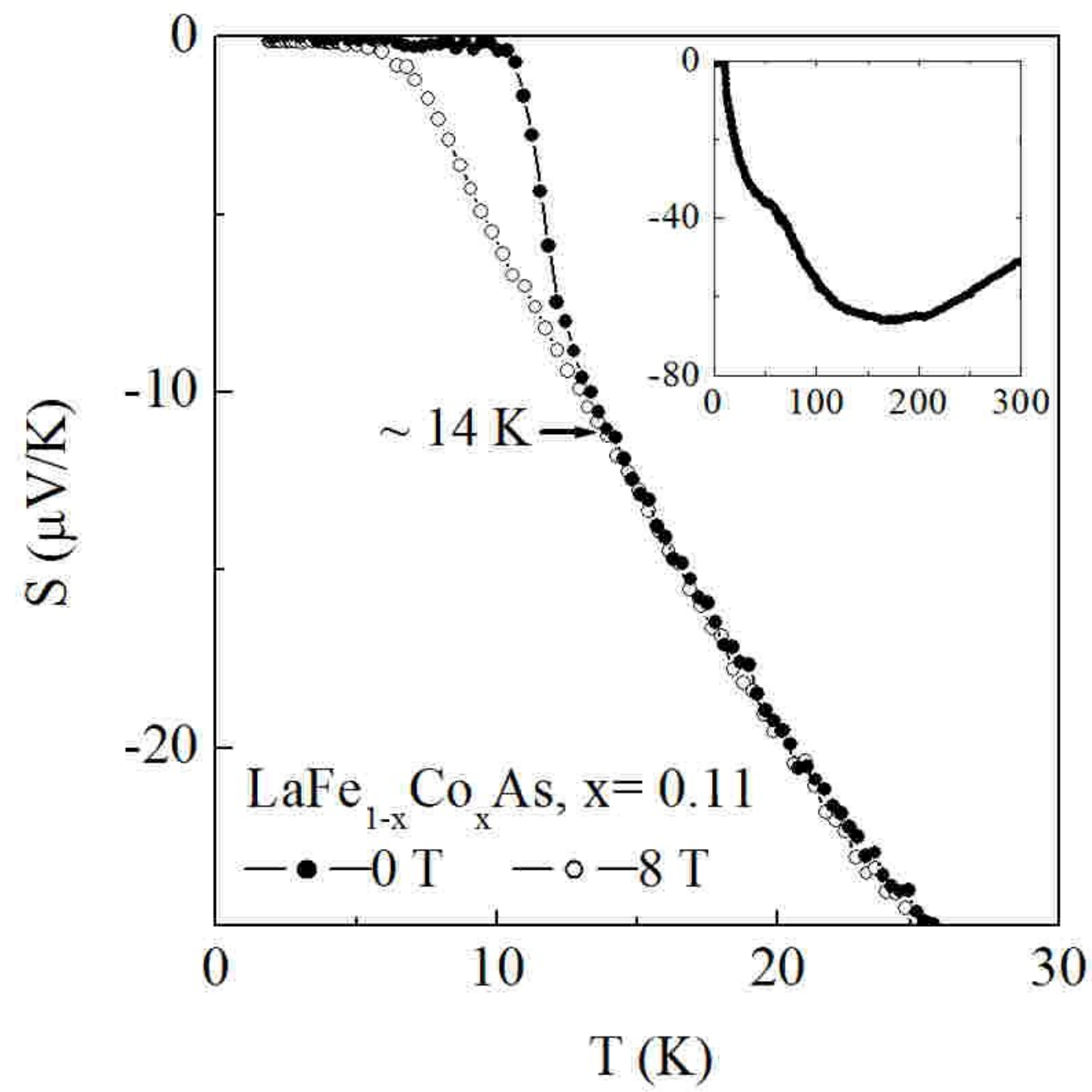

Fig. 9. Temperature dependence of Seebeck coefficient for $LaFe_{1-x}Co_xAsO$ with x = 0.11, in applied fields of 0 and 8 T. Inset is zero field result, up to room temperature.

Table 1: Crystallographic data of LaFeAsO

| T (K) | 4 | 300 |
|---|---|---|
| Space group | *Cmma* | *P*4/*nmm* |
| *a* (Å) | 5.6823(2) | 4.0345(1) |
| *b* (Å) | 5.7103(2) | = *a* |
| *c* (Å) | 8.7117(4) | 8.7387(4) |
| $V$ (Å$^3$) | 282.67(2) | 142.24(1) |
| $Z$ | 4 | 2 |
| *d*-range (Å) | 0.8 - 10.6 | |
| $wR_p$ | 3.17 | 3.20 |
| $\chi^2$ | 7.14 | 5.92 |
| *Atomic parameters:* | | |
| La | 4*g* (0,¼,*z*), z= 0.1424(3)<br>$U_{iso}$ = 0.0025(8) Å$^2$ | 2*c* (¼,¼,*z*), z= 0.1420(3)<br>$U_{iso}$= 0.0066(11) Å$^2$ |
| Fe | 4*b* (¼,0,½), $U_{iso}$= 0.0011(6) | 2*b* (¾,½,*z*), $U_{iso}$= 0.0055(8) Å$^2$ |
| As | 4*g* (0,¼,*z*), z= 0.6501(4)<br>$U_{iso}$= 0.0009(10) Å$^2$ | 2*c* (¼,¼,*z*), z= 0.6498(4),<br>$U_{iso}$= 0.0045(12) Å$^2$ |
| O | 4*a* (¼,0,0), $U_{iso}$= 0.0049(11) Å$^2$ | 2*a* (¾,¼,0), $U_{iso}$= 0.0096(13) Å$^2$ |

Table 2: Crystallographic data of $LaFe_{1-x}Co_xAsO$: x = 0.11

| T (K) | 4 | 300 |
|---|---|---|
| Space group | *P*4/*nmm* | *P*4/*nmm* |
| *a* (Å) | 4.02771(1) | 4.0351(1) |
| *b* (Å) | = *a* | = *a* |
| *c* (Å) | 8.6860(3) | 8.7132(3) |
| $V$ (Å$^3$) | 140.908(6) | 141.871(11) |
| $Z$ | 2 | 2 |
| *d*-range (Å) | 0.8 - 10.6 | |
| $wR_p$ | 3.38 | 3.61 |
| $\chi^2$ | 3.94 | 4.25 |
| *Atomic parameters:* | | |
| La | 2*c* (¼,¼,*z*), z= 0.1412(3)<br>$U_{iso}$= 0.0038(8) Å$^2$ | 2*c* (¼,¼,*z*), z= 0.1412(3)<br>$U_{iso}$= 0.0049(9) Å$^2$ |
| Fe/Co | 2*b* (¾,½,*z*), $U_{iso}$= 0.0015(8) Å$^2$ | 2*b* (¾,½,½),$U_{iso}$= 0.0087(11) Å$^2$ |
| As | 2*c* (¼,¼,*z*), z= 0.6502(3),<br>$U_{iso}$ = 0.0012(9) Å$^2$ | 2*c* (¼,¼,*z*), z= 0.6505(4),<br>$U_{iso}$ = 0.0054(12) Å$^2$ |
| O | 2*a* (¾,¼,0), $U_{iso}$= 0.0057(11) Å$^2$ | 2*a* (¾,¼,0), $U_{iso}$= 0.0057(14) Å$^2$ |